\documentclass{article}
\usepackage{graphicx} 
\usepackage{amsmath,amsthm,amsfonts}
\usepackage[margin=1in]{geometry}
\usepackage{algorithm}
\usepackage{algpseudocode}
\usepackage{color}

\usepackage{xcolor}

\title{On Unbiased Low-Rank Approximation with Minimum Distortion}
\author{Leighton Pate Barnes, Stephen Cameron, and Benjamin Howard \\ {\small Center for Communications Research -- Princeton}}
\date{May 2025}

\begin{document}

\maketitle

\begin{abstract}
We describe an algorithm for sampling a low-rank random matrix $Q$ that best approximates a fixed target matrix $P\in\mathbb{C}^{n\times m}$ in the following sense: $Q$ is unbiased, i.e., $\mathbb{E}[Q] = P$; $\mathsf{rank}(Q)\leq r$; and $Q$ minimizes the expected Frobenius norm error $\mathbb{E}\|P-Q\|_F^2$. Our algorithm mirrors the solution to the efficient unbiased sparsification problem for vectors, except applied to the singular components of the matrix $P$. Optimality is proven by showing that our algorithm matches the error from an existing lower bound.
\end{abstract}

\section{Introduction}
Suppose that $P$ is any complex-valued $n$ by $m$ matrix, and we would like to approximate it with a randomly sampled matrix $Q$ (also in $\mathbb{C}^{n\times m}$) that is low rank. We require that $\mathsf{rank}(Q)\leq r$ for any realization of $Q$, and that the sampling is unbiased in the sense that $\mathbb{E}[Q]=P$ entrywise. In this note, we show that a simple sampling scheme inspired by the problem of \emph{efficient unbiased sparsification} \cite{barnes2024efficient}, if applied to the singular components of $P$, produces a random matrix $Q$ that minimizes the expected distortion \begin{equation}\mathbb{E}\|P-Q\|_F^2 \label{eq:frob} \end{equation}
where the expectation is over the random sampling and the norm is the usual Frobenius norm.

The problem of unbiased low-rank approximation has recently been considered in the machine learning literature, and in particular the authors \cite{benzing2019optimal} show that a duality-based lower bound on the expected distortion can be matched by a complicated sampling procedure. In \cite{benzing2019optimal}, their procedure is explicitly defined for the co-rank one scenario $r+1 = \mathsf{rank}(P)$, and for arbitrary $r$ they point out that an algorithm can be written down inductively in $r$ (see the section titled \emph{Proof of Lemma A.6} in the supplementary material). Our main contribution in the present note is to show that a simple sparsification procedure applied to the singular components of $P$ achieves this lower bound for any $r$, and therefore gives a simpler and more direct sampling procedure for producing optimal, unbiased, low-rank approximations.

Our sampling procedure mirrors the optimal way to sample a sparse approximation of a dense vector that is described in \cite{barnes2024efficient}. One main difference is that the present work is basis-independent; we apply this procedure to the singular components of the matrix $P$, and it turns out to be optimal among all rank-$r$ approximations. It is not obvious that we can work only in the basis described by the singular components of $P$, and indeed, the sampling procedure from \cite{benzing2019optimal} does not work this way. In other words, suppose $P$ were diagonal. Then there exist optimal, unbiased, low-rank approximations that are both strictly diagonal (as in our sampling procedure), and not strictly diagonal (as in \cite{benzing2019optimal}).  This is true even when there are no repeated singular values.

To give a concrete example, suppose that we want a rank 1 unbiased approximation of $$P = \begin{bmatrix} 4 & 0 \\ 0 & 1 \end{bmatrix} \; .$$ Our Algorithm \ref{fig1} introduced below generates a random variable $Q$ which takes values 
\begin{equation*}
Q = \begin{cases}\begin{bmatrix} 5 & 0 \\ 0 & 0\end{bmatrix} & \text{, with probability } 4/5 \\ \\
\begin{bmatrix} 0 & 0 \\ 0 & 5\end{bmatrix} & \text{, with probability } 1/5 \; ,
\end{cases}
\end{equation*}
whereas the algorithm in \cite{benzing2019optimal} generates the random variable $$Q' = \begin{bmatrix} 4 & \pm 2 \\ \pm 2 & 1 \end{bmatrix}$$ each with equal probability 1/2.  Both $Q,Q'$ achieve the minimal distortion $\mathbb{E}\|P-Q\|_F^2  = \mathbb{E}\|P-Q'\|_F^2 = 8$, as does the mixture $Q_t$ which takes the values $Q_t = Q$ with probability $t$ and $Q_t = Q'$ with probability $1-t$ for any $0\leq t \leq 1$.  This variety of optimal solutions is in stark contrast to the rigid description of optimizers for the vector valued case \cite{barnes2024efficient}, and suggests there is a rich space of possible optimal unbiased approximations to explore.

In the work \cite{barnes2024efficient}, the vector sparsification procedure is shown to be optimal for a wide range of divergence measures including both the KL divergence and the squared Euclidean distance. Unknown to the authors of \cite{barnes2024efficient} at the time of publication, this procedure was already known to be optimal in the special case of squared Euclidean distance as described in \cite{fearnhead2003line}. However, \cite{barnes2024efficient} shows that it holds in much more generality and is simultaneously optimal for large classes of divergence measures.

Most other works that have considered unbiased approximation of matrices are concerned more with computational constraints, rather than getting the exact approximation that minimizes some error \cite{frieze2004fast,vogels2019powersgd}; consider a noisy low-rank recovery problem \cite{carlsson2022unbiased}; or only focus on the rank one case \cite{ollivier2015training,tallec2017unbiased}. Our unbiased low-rank approximation procedure is conceptually similar to the spectral version of \emph{atomo} \cite{wang2018atomo} from the federated learning literature. However, we consider a ``hard'' constraint where each realization of $Q$ has at most rank $r$, instead of their ``soft'' constraint which only requires that the expected number of singular components be at most $r$, and they do not establish the optimality of spectral \emph{atomo} with respect to Frobenius norm error. As far as we know, the optimality of these types of sampling procedures in the sense of \eqref{eq:frob} is not previously known in the literature.

\section{The Sampling Procedure}
First note that by taking a singular value decomposition, we can restrict our attention to approximating a target matrix $\Lambda$ that is real, non-negative, and diagonal. This is because the Frobenius norm is invariant to unitary transformations, and if $P=U\Lambda V^*$ for unitary $U,V$ then
$$\|P-Q\|_F = \|\Lambda - U^*QV\|_F$$ and $Q' = U^*QV$ is an unbiased rank-$r$ approximation of $\Lambda$. Without loss of generality we assume $$\Lambda = \mathsf{diag}(d_1,\ldots,d_N,0,\ldots,0)$$ where $d_1\geq d_2\geq \ldots \geq d_N$ and $N=\mathsf{rank}(P)$.

Without yet knowing that it might be optimal, suppose our goal was to construct an unbiased rank-$r$ approximation of $\Lambda$ that is also diagonal. A reasonable idea might be to include component $i$ as a nonzero component of $Q'$ with probability proportional to $d_i$. Letting $C_1$ be this constant of proportionality, we would have
$$r = \sum_{i=1}^N p_i = \sum_{i=1}^N C_1d_i $$
and
$$C_1 = \frac{r}{\sum_{i=1}^N d_i} \; . $$
In particular, the largest component $d_1$ would be included with probability
$$p_1 = \frac{rd_1}{\sum_{i=1}^N d_i} \; .$$
If this probability is greater than one, then our original idea needs modification. In this case, let's simply include the first component in \emph{every} realization of $Q'$, and repeat the procedure on the remaining components. This results in a number of components $k$ (called the \emph{heavy} components \cite{barnes2024efficient}), which is the smallest number such that
$$\frac{(r-k)d_{k+1}}{\sum_{i=k+1}^N d_i} < 1  \iff (r-k) d_{k+1} < \sum_{i=k+1}^N d_i\; ,$$
and these components are automatically included in every realization of the diagonal $Q'$. This notion of the first $k$ indices being heavy (where $k$ is potentially zero) is well-defined in the sense that
$$(r-k')d_{k'+1} < \sum_{i=k'+1}^N d_i \implies (r-(k'+1))d_{k'+2} < \sum_{i=k'+2}^N d_i \; .$$

Of the remaining $N-k$ components, sample a subset $I$ of $r-k$ elements of the indices $\{k+1,\ldots,N\}$ with inclusion probabilities proportional to their $d_i$ values. There are many ways to do this sampling, and we describe one simple way in the following paragraph (see also \cite{barnes2024efficient,tille2006sampling}). Let's then set the resulting $Q'=\mathsf{diag}(Q'_1,\ldots,Q'_N,0,\ldots,0)$ by the components
\begin{equation} \label{eq:sparsecomps}
Q_i' = \begin{cases} d_i & \; , \; i=1,\ldots,k \\ c = \frac{\sum_{j=k+1}^N d_j}{r-k} & \; , \; i\in I \\ 0 & \; , \; \mathsf{otherwise} \; . \end{cases}
\end{equation}
The matrix $Q'$ is an unbiased, rank-$r$ approximation of $\Lambda$ by construction.

As promised, one way to sample the index set $I$ is as follows. Divide the interval $[0,r-k]$ up into $N-k$ disjoint segments, with segment $i$ (for $i=k+1,\ldots,N$) having length proportional to $d_i$. Then randomly generate a single number $S\sim\mathsf{unif}[0,1]$, and if $S+j$ is in segment $i$ for some $j=0,1,\ldots,r-k-1$, then include the index $i$ in $I$. Otherwise, do not include that index $i$ in $I$. It may be desirable to randomly permute the order the segments, but this is not required to achieve the correct inclusion probabilties.

One can verify this gives the desired inclusion probabilities by zooming in on a single, unit-length subinterval $[z,z+1],z\in\mathbb{Z}$ of $[0,r-k]$. If this subinterval entirely contains segment $i$, then segment $i$ will be included with probability equal to its length (which is proportional do $d_i$). If part of segment $i$ lies in an adjacent subinterval, then the sum of the lengths of the two parts will be the inclusion probability (this uses the fact that no segment can have length more than one, so the event that both parts are chosen cannot happen). No segment can intersect more than two subintervals with nonzero measure, since their lengths are bounded by one.

\begin{figure}
    \begin{center}
    \includegraphics[scale=0.4]{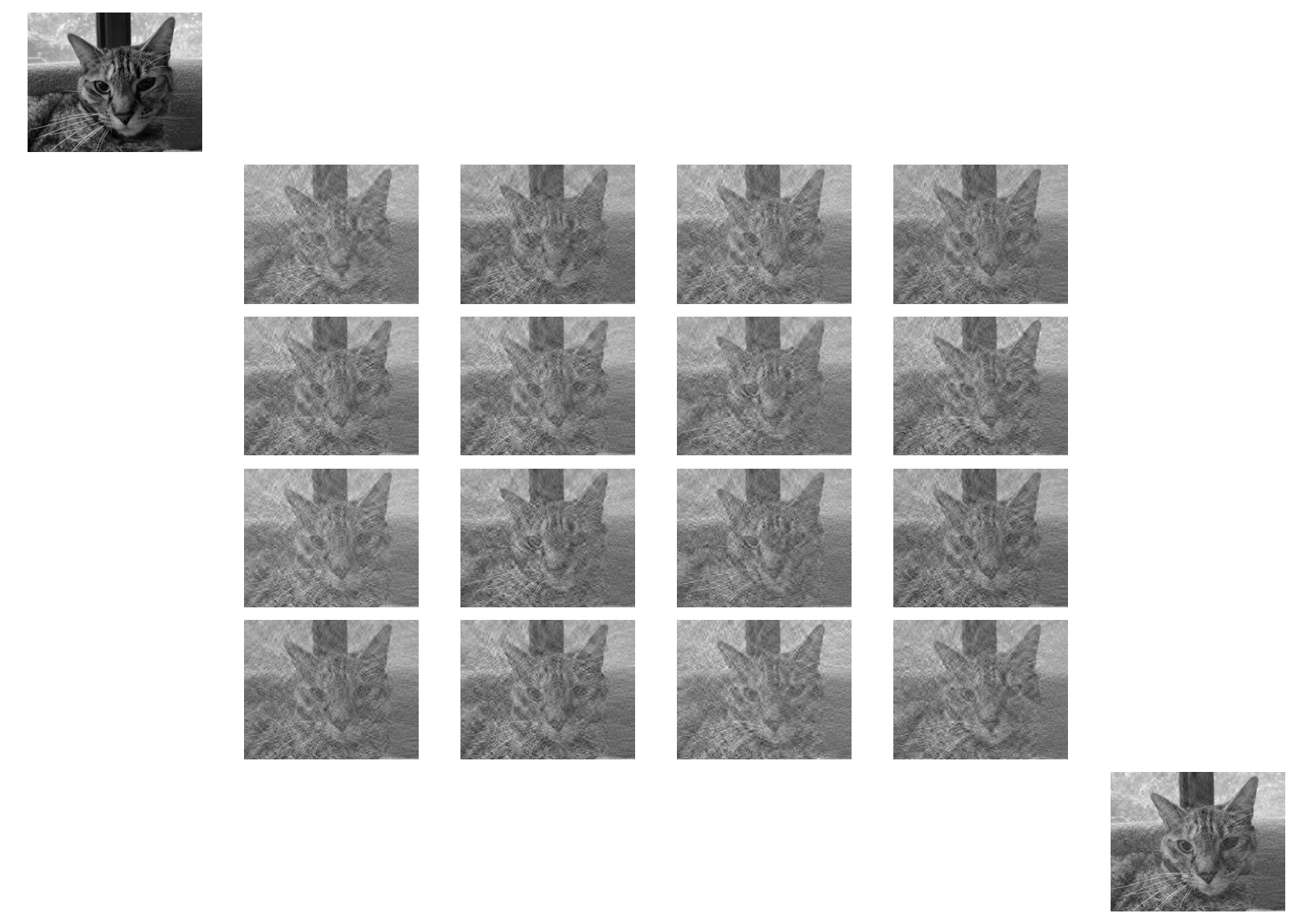}
    \caption{Upper-left is an image of the third author's pet cat ``Nutmeg'', of rank $288$, and the lower-right is the average 
    of 16 unbiased low-rank approximations, all instances of Algorithm \ref{fig1} with rank $r=30$. 
    The images in the middle are the unbiased approximations.}\label{fig:cats}
    \end{center}
\end{figure}

\begin{figure}
    \begin{center}
    \includegraphics[scale=0.4]{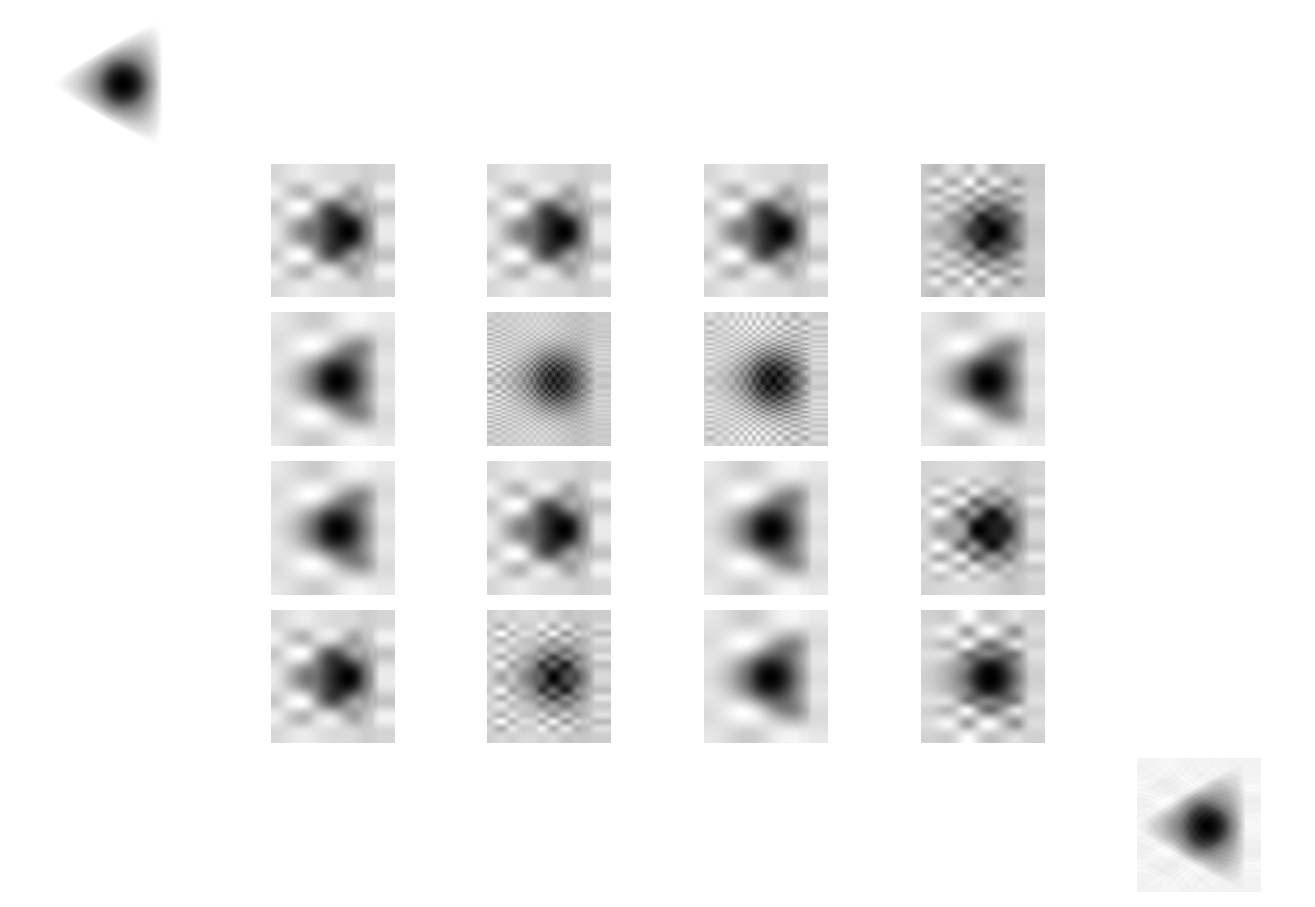}
    \caption{Upper-left is an image of rank $\approx 700$, and the lower-right is the average 
    of 16 unbiased low-rank approximations, all instances of Algorithm \ref{fig1} with rank $r=3$. 
    The images in the middle are the unbiased approximations.}\label{fig:triangles}
    \end{center}
\end{figure}

Putting this all together, we summarize the unbiased low-rank approximation procedure in Algorithm \ref{fig1} below. An illustration of the algorithm applied to grayscale images (treated as real-valued matrices) is shown in Figures \ref{fig:cats} and \ref{fig:triangles}. The runtime of the algorithm is dominated by the cost of computing the singular value decomposition of the matrix $P$ at the beginning.

\begin{algorithm}
\caption{Unbiased Low-Rank Approximation with Minimum Distortion}
\label{fig1}
\begin{algorithmic}[1]
\item {\bf given} matrix $P\in\mathbb{C}^{n\times m}$, rank constraint $r$
\item {\bf compute SVD} $P=U\Lambda V^*$, $\Lambda = \mathsf{diag}(d_1,\ldots,d_N,0,\ldots,0)$
\For{k'=0,\ldots,N-1}
\If {$(r-k')d_{k'+1} < \sum_{i=k'+1}^N d_i$} \State $k\leftarrow k'$
\State break
\EndIf
\EndFor
\item $S\leftarrow\mathsf{unif}[0,1]$
\item $I\leftarrow \{\}$
\item $\mathsf{length} \leftarrow 0$
\item $C \leftarrow \frac{r-k}{\sum_{i=k+1}^Nd_i}$
\For{i=k+1,\ldots,N}
\State $\mathsf{length} \leftarrow \mathsf{length} + Cd_i$
\If {$\mathsf{length} \geq S$}
\State $I\leftarrow I \cup \{i\}$
\State $S \leftarrow S+1$
\EndIf
\EndFor
\item $Q'=\mathsf{diag}(Q'_1,\ldots,Q'_N,0,\ldots,0)$ where $
Q_i' = \begin{cases} d_i & \; , \; i=1,\ldots,k \\ c = \frac{1}{C} = \frac{\sum_{j=k+1}^N d_j}{r-k} & \; , \; i\in I \\ 0 & \; , \; \mathsf{otherwise} \end{cases}
$
\item {\bf return} $Q = UQ'V^*$
\end{algorithmic}
\end{algorithm}

\section{Matching Lower Bound}
In this section, we show that the lower bound from the supplementary material in \cite{benzing2019optimal} matches the expected error achieved by Algorithm \ref{fig1}. This shows that Algorithm \ref{fig1} produces unbiased low-rank approximations that are optimal in the sense that they minimize \eqref{eq:frob}.

For any fixed matrix $B$, we have
\begin{align}
\mathbb{E}\|Q'-\Lambda\|_F^2 & = \mathbb{E}\|Q'-(\Lambda-B)\|_F^2 - \|B\|_F^2 \label{eq:step1}\\
& \geq \min_{X\in\mathbb{C}^{n\times m}, \, \mathsf{rank}(X)\leq r} \|X-(\Lambda-B)\|_F^2 - \|B\|_F^2 \label{eq:step2} \; .
\end{align}
Equation \eqref{eq:step1} follows by the unbiasedness of $Q'$ so that $\mathbb{E}[B^*(Q'-\Lambda)] = 0$. In equation \eqref{eq:step2}, we lower bound the expected error by the smallest error that can be achieved, with any fixed low-rank matrix $X$, between $\Lambda-B$ and $X$. In particular, we pick $B$ so that
\begin{equation} \label{eq:defineB}
\Lambda - B = \mathsf{diag}(d_1,\ldots,d_k,\underbrace{c,\ldots,c}_{N-k},0,\ldots,0)
\end{equation}
where $c$ is defined as in \eqref{eq:sparsecomps}.

A key property that is needed is that $d_k \geq c$ so that the diagonal elements in \eqref{eq:defineB} are non-increasing. This follows because by definition the index $k$ is heavy, so
$$ (r-(k-1))d_k \geq \sum_{j=k}^Nd_j \implies d_k \geq \frac{\sum_{j=k+1}^Nd_j}{r-k} = c\; .$$
Using this property and the Eckart-Young-Mirsky Theorem, it is clear that the expression in \eqref{eq:step2} is minimized exactly by the first $r$ diagonal components of \eqref{eq:defineB}. Furthermore, if $Q'$ is defined as in \eqref{eq:sparsecomps}, then the error in \eqref{eq:step1} is the same as in \eqref{eq:step2} for \emph{every} realization of $Q'$. Therefore, for our choice of $Q'$, the inequality in \eqref{eq:step2} is actually equality, and we exactly match the minimum possible expected error.

\bibliography{main.bib}
\bibliographystyle{IEEEtran}

\end{document}